\documentclass[aps,pra,reprint,superscriptaddress,showpacs]{revtex4-1}
\usepackage{graphicx,amsmath,amssymb,ulem}
\usepackage[T1]{fontenc}
\usepackage[english]{babel}
\usepackage{float}

\usepackage{url}

\usepackage{soul}
\usepackage[dvipsnames]{xcolor}
\definecolor{darkred}{HTML}{A80000}

\DeclareMathOperator{\sinc}{sinc}

\begin{document}

\title{The multimode four-photon Hong-Ou-Mandel interference}

\author{A.~Ferreri}
\affiliation{Department of Physics, Paderborn University,
Warburger Strasse 100, D-33098 Paderborn, Germany}
\author{V.~Ansari}
\affiliation{Department of Physics, Paderborn University,
Warburger Strasse 100, D-33098 Paderborn, Germany}
\author{C.~Silberhorn}
\affiliation{Department of Physics, Paderborn University,
Warburger Strasse 100, D-33098 Paderborn, Germany}
\author{P.~R.~Sharapova}
\affiliation{Department of Physics, Paderborn University,
Warburger Strasse 100, D-33098 Paderborn, Germany}

\begin{abstract}
The two-photon Hong-Ou-Mandel (HOM) interference is a pure quantum effect which indicates the degree of indistinguishability of photons.  The four-photon HOM interference exhibits richer dynamics in comparison to the two-photon interference and simultaneously is more sensitive to the input photon states. We demonstrate theoretically and experimentally an explicit dependency of the four-photon interference to the number of temporal modes, created in the process of parametric down-conversion. Moreover, we exploit the splitting ratio of the beam splitter to manipulate the interference between bunching and antibunching. Our results reveal that the temporal mode structure (multimodeness) of the quantum states shapes many-particle interference.

\end{abstract}
\maketitle

\section{Introduction}
A useful tool to investigate the degree of indistinguishability of photons in quantum optics is the  Hong-Ou-Mandel (HOM) interference \cite{hong1987measurement}. Beyond the interest on the peculiar quantum interference itself, HOM interference is extremely important in quantum information, e.g. for the Bell-state measurement \cite{mattle1996dense}, it is used to testify the non-locality of entangled system \cite{torgerson1995experimental} and in quantum lithography \cite{nagasako2001nonclassical, boto2000quantum}. Typically this kind of interference is studied by involving two photons in free space, though in the last years integrated quantum devices are also utilized because of their helpful features which allow to manipulate the interference process  \cite{PhysRevA.96.043857, lim2005generalized}.

However, the multiphoton interference which involves more than two photons attracts a lot of attention nowadays \cite{ tichy2012many, PhysRevA.60.593, PhysRevA.87.062106, ou1999observation} because it is an essential and  indispensable tool for boson sampling \cite{bentivegna2015experimental} and machine learning \cite{shen2017deep, hughes2018training, tait2017neuromorphic, flamini2019visual} which are the first steps of future quantum computing. Moreover, the multiphoton interference  allows to achieve  a high-dimensional entanglement \cite{walther2004broglie}, overcome the standard quantum limit in interferometry \cite{nagata2007beating} and create high dimensional NOON states \cite{afek2010high}. In addition, the multiphoton interference can highlight and solve the fundamental question about the quantum-to-classical transition. In this way, non-monotonic character of the photon interference with increasing the number of photons was studied in  \cite{ra2013nonmonotonic}. Behaviour of the four-photon interference with increasing a pump power was investigated in \cite{PhysRevA.77.053822}. The multiphoton interference is directly connected with the multiparticle indistinguishability and the collective phase of photons  \cite{PhysRevLett.118.153603,  ou2005distinguishing}, theoretical description of the multiphoton interference with transition matrix is presented in  \cite{PhysRevX.5.041015}.

In this work we investigate the four-photon interference in relation to the spectral-temporal properties of the photons generated via parametric down-conversion. We show that the number of temporal modes drastically influences the interference pattern,  observing a raising of HOM dip/peak in the coincidence probability. Our analysis takes into account also different beam splitter  parameters, allowing us to control the interference visibility. We show that the temporal mode structure, particularly the amount of multimodeness, of the photon source plays a crucial role in the multiphoton interference in contrast to the two-photon interference \cite{PhysRevA.96.043857}.  

\section{Theoretical model}
The type-II parametric down-conversion (PDC) process produces pairs of photons related by both frequency and polarization entanglement \cite{weinfurter2001four, howell2002experimental}.  The Hamiltonian of the type-II PDC process can be written in terms of the joint spectral amplitude (JSA)$F(\omega_s, \omega_i)$ \cite{quesada2014effects}:
\begin{equation}
H= \Gamma \int \mathsf{d}\omega_s \mathsf{d}\omega_i F(\omega_s, \omega_i) a_H^\dagger(\omega_s)a_V^\dagger(\omega_i) + \mathsf{H}. \mathsf{c.},
\end{equation}
where the indices "s" and "i" indicate the "signal" and "idler" photons respectively, $a_{s,(i)}$ and $a_{s,(i)}^\dagger$ are the annihilation and creation operators of the signal (idler) photons, the coupling constant $\Gamma$ determines the strength of interaction, $H$ and $V$ label the horizontal and vertical polarization of the photons.
In a periodically poled medium with the poling period $\Lambda$ the JSA can be written in the form
\begin{equation}
F(\omega_s, \omega_i)=e^{-\frac{(\omega_s+\omega_i-\omega_p)}{2 \Omega^2}}\sinc\bigg(\frac{L}{2}\Delta k\bigg)e^{i \frac{L}{2}\Delta k},
\end{equation}
where $\Omega$ is the pump spectral bandwidth, $L$ is the length of the PDC section, $\omega_p$ is the pump center frequency, $\Delta k=k_p(\omega_p)-k_s(\omega_s)-k_i(\omega_i)+\frac{2\pi}{\Lambda}$ is the phase matching condition which determines the momentum conservation of the process.

The four-photon state generated in the PDC process can be described by using the second-order of the perturbation theory. Neglecting the time-ordering effect the generated four-photon state is \cite{ansari2014probing, riedmatten2004two, shi2006four}
\begin{equation}
\begin{aligned}
|\psi_{4 ph}\rangle = & \frac{1}{2}\bigg( \int_0^t H(t')dt'\bigg)^2 \mid 0 \rangle \\
= & \frac{\xi^2 }{2}\int^{+\infty}_{-\infty}\mathsf{d}\omega_s\mathsf{d}\omega_i  F(\omega_s,\omega_i)
a_H^\dagger(\omega_s)a_V^\dagger(\omega_i)\times \\ 
&\int^{+\infty}_{-\infty} \mathsf{d}\tilde\omega_s \mathsf{d}\tilde\omega_i F(\tilde\omega_s,\tilde\omega_i) a_H^\dagger(\tilde\omega_s)a_V^\dagger(\tilde\omega_i)|0\rangle,
\end{aligned}
\end{equation} 
where $\omega_s, \omega_i, \tilde{\omega}_s, \tilde{\omega}_i$ are the frequencies of the four generated photons and $\xi = \Gamma t$, where $t$ is the time of the interaction process.

The JSA strongly depends on dispersion properties of the non-linear material where PDC takes place \cite{kato2002sellmeier}. For our investigation we choose a ppKTP waveguide since this material finds flexibility to the number of Schmidt modes: a ppKTP waveguide is able to provide a quasi-single-mode PDC state, as well as the strongly multimode regime by varying the pulse duration only. Such property is more difficult to obtain in other materials (LiNbO$_3$, BBO) without filtering.

\begin{figure}[t]
\includegraphics[width=0.5%
\textwidth]{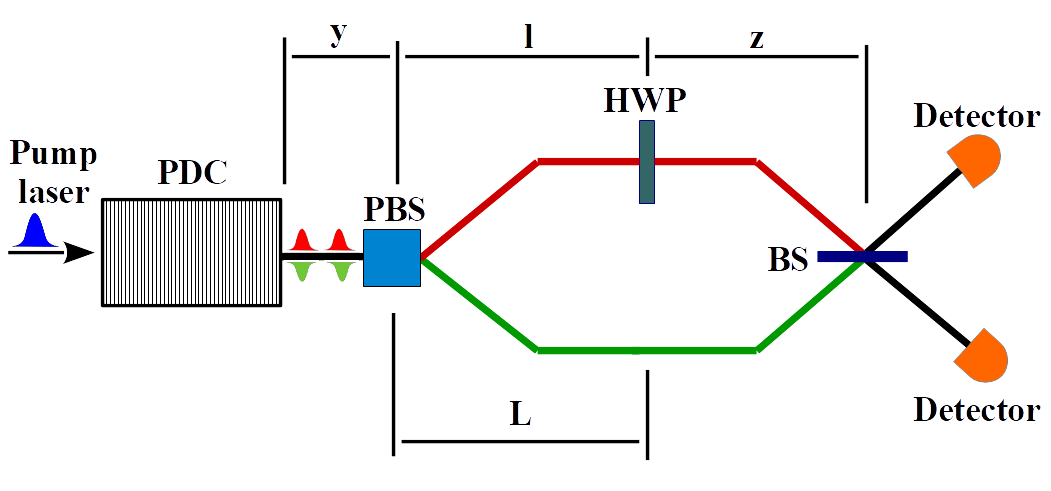}
\caption{Schematic setup. The type-II PDC process generates two signal-idler pairs of photons. After PBS, two horizontally-polarized photons are routed to the upper channel (red line), while two vertically-polarized photons, are routed to the lower channel (green line). A PC located in the upper channel converts the horizontally-polarized photons into the vertically-polarized. An additional path increment $L=l+\Delta l$ in the lower channel allows to compensate the time delay between the signal and idler photons. Then four vertically-polarized photons cross the BS at the same time, the HOM interference occurs. The photons are detected after the BS. }
\label{chip}
\end{figure}

To observe the HOM interference, the incoming photons into a beam splitter (BS) must be fully indistinguishable in all degrees of freedom: must have the same polarization and cross the BS without any time delay. To satisfy both conditions, we consider the following setup schematically sketched in Fig.\ref{chip}.
In Fig.\ref{chip} the generated photons inside the PDC section  are separated then by a polarization beam splitter (PBS) in two different spatial channels. The horizontally-polarized photons is routed in the upper channel, and the vertically-polarized photons in the lower channel. A polarization converter (PC) located in the upper channel converts the horizontally-polarized photons into the vertically-polarized. Due to the different group velocities of the horizontally- and vertically-polarized photons inside the nonlinear crystal, a time delay between the signal and idler photons is present already after the PDC section. To compensate this time delay we create an additional path increment $\Delta l$ in the lower channel. In the end of the setup the four vertically-polarized photons cross the BS at the same time and interfere. The interference pattern is measured at the output ports of the BS using photon number resolving detectors.

Mathematically, all these transformations can be expressed as a unitary matrices which act on the initial four-photon state. The final unitary transformation can therefore be written as a product of unitary matrices:
\begin{equation}
U_{tot} = BS\times FP_3\times PC\times FP_2\times PBS\times FP_1,
\end{equation}
where $BS$, $PC$ and $PBS$ are the matrices of the concerning elements and the matrices $FP_i$ describe a free propagation of the light between optical elements. The output state after the all transformations can be obtained after the action of the total matrix $U_{tot}$ on the initial four-photon state. In the basis $\{a_{1H}^\dagger, a_{1V}^\dagger, a_{2H}^\dagger, a_{2V}^\dagger \}$ , where the indexes 1, 2 correspond to the upper and lower channels, the output state can be represented by
\begin{widetext}
\begin{equation}\begin{split}
|\psi_{out}\rangle= \frac{\xi^2 }{2}\int \mathsf{d}\omega_s \mathsf{d}\omega_i  \mathsf{d}\tilde\omega_s \mathsf{d}\tilde\omega_i F(\omega_s,\omega_i)F(\tilde\omega_s,\tilde\omega_i) \times
U_0(\omega_s)U_{tot}^\dagger(\omega_s)
\begin{pmatrix}
a_{1H}^\dagger(\omega_s) \\ a_{1V}^\dagger(\omega_s) \\ a_{2H}^\dagger(\omega_s) \\ a_{2V}^\dagger(\omega_s)
\end{pmatrix} 
 \otimes U_0(\omega_i)U_{tot}^\dagger(\omega_i)
\begin{pmatrix}
a_{1H}^\dagger(\omega_i) \\ a_{1V}^\dagger(\omega_i) \\ a_{2H}^\dagger(\omega_i) \\ a_{2V}^\dagger(\omega_i)
\end{pmatrix}  \ \ \ \ \ \ \\
 \otimes U_0(\tilde\omega_s)U_{tot}^\dagger(\tilde\omega_s)
\begin{pmatrix}
a_{1H}^\dagger(\tilde\omega_s) \\ a_{1V}^\dagger(\tilde\omega_s) \\ a_{2H}^\dagger(\tilde\omega_s) \\ a_{2V}^\dagger(\tilde\omega_s)
\end{pmatrix} 
 \otimes U_0(\tilde\omega_i)U_{tot}^\dagger(\tilde\omega_i)
\begin{pmatrix}
a_{1H}^\dagger(\tilde\omega_i) \\ a_{1V}^\dagger(\tilde\omega_i) \\ a_{2H}^\dagger(\tilde\omega_i) \\ a_{2V}^\dagger(\tilde\omega_i)
\end{pmatrix} 
|0\rangle , \ \ 
\end{split}
\label{psi_out}
\end{equation}
\end{widetext}
where $U_0(\omega_s)$ and $U_0(\omega_i)$ are the initial condition matrices \cite{sharapova2017toolbox}. The output state Eq.(\ref{psi_out}) can be modified and written in the basis $|m,n \rangle$ with the probability amplitudes $C_{m,n}(\Delta l, \tau , L)$,  where $m$ indicates the number of photons in the upper channel, $n$ - the number of photons in the lower channel: 
\begin{equation}\begin{split}
|\psi_{out}\rangle=\int \mathsf{d}\omega_s \mathsf{d}\omega_i \mathsf{d}\tilde\omega_s \mathsf{d}\tilde\omega_i (C_{22}(\Delta l,\tau,  L)|2,2 \rangle+ \ \ \ \ \ \ \\\
 C_{31}(\Delta l,\tau,  L)(|3,1 \rangle+|1,3 \rangle)+ C_{40}(\Delta l,\tau,  L)(| 4,0 \rangle+| 0,4 \rangle)).
\end{split}
\label{out}
\end{equation}

By using the final state Eq. (\ref{out}), we can calculate the expectation values of the  simultaneous positive-operator valued measures (POVM) which corresponds to the measuring the coincidence probability at the detectors :
\begin{eqnarray}
P_{22}(\Delta l,\tau, L)=\int \mathsf{d}\omega_b \mathsf{d}\omega_c \mathsf{d}\tilde\omega_b \mathsf{d}\tilde\omega_c \nonumber \\
 |\langle 0| \frac{1}{\sqrt{2!}\sqrt{2!}}d_1 (\omega_b)d_2(\omega_c)d_1(\tilde\omega_b)d_2(\tilde\omega_c)|\tilde{\psi}_{out}\rangle|^2
\label{P22}
\end{eqnarray}
\begin{eqnarray}
P_{31}(\Delta l,\tau, L)=\int \mathsf{d}\omega_b \mathsf{d}\omega_c \mathsf{d}\tilde\omega_b \mathsf{d}\tilde\omega_c \nonumber \\ 
|\langle 0| \frac{1}{\sqrt{3!}}d_1 (\omega_b)d_1(\omega_c)d_1(\tilde\omega_b)d_2(\tilde\omega_c)|\tilde{\psi}_{out}\rangle|^2
\label{P31}
\end{eqnarray}
\begin{eqnarray}
P_{40}(\Delta l,\tau, L)=\int \mathsf{d}\omega_b \mathsf{d}\omega_c \mathsf{d}\tilde\omega_b \mathsf{d}\tilde\omega_c \nonumber \\
|\langle 0| \frac{1}{\sqrt{4!}}d_1 (\omega_b)d_1(\omega_c)d_1(\tilde\omega_b)d_1(\tilde\omega_c)|\tilde{\psi}_{out}\rangle|^2,
\label{P40}
\end{eqnarray}
where  $d_1$ and $d_2$ are the annihilation operators of the detectors placed in the upper and the lower channels respectively, $\omega_b,\omega_c, \tilde{\omega}_b, \tilde{\omega}_c $ are the frequencies of the detectors, $|\tilde{\psi}_{out}\rangle=|\psi_{out}\rangle / \langle \psi_{out}|\psi_{out}\rangle$ is the normalized output state.
The Eq. (\ref{P22},\ref{P31},\ref{P40}) present the probability $P_{mn}$ to detect $m$ photons in the upper channel and $n$ photons in the lower channel respectively.

\section{Results and discussion}

\subsection{Experiment}
\begin{figure}[t]
\includegraphics[width=1\linewidth]{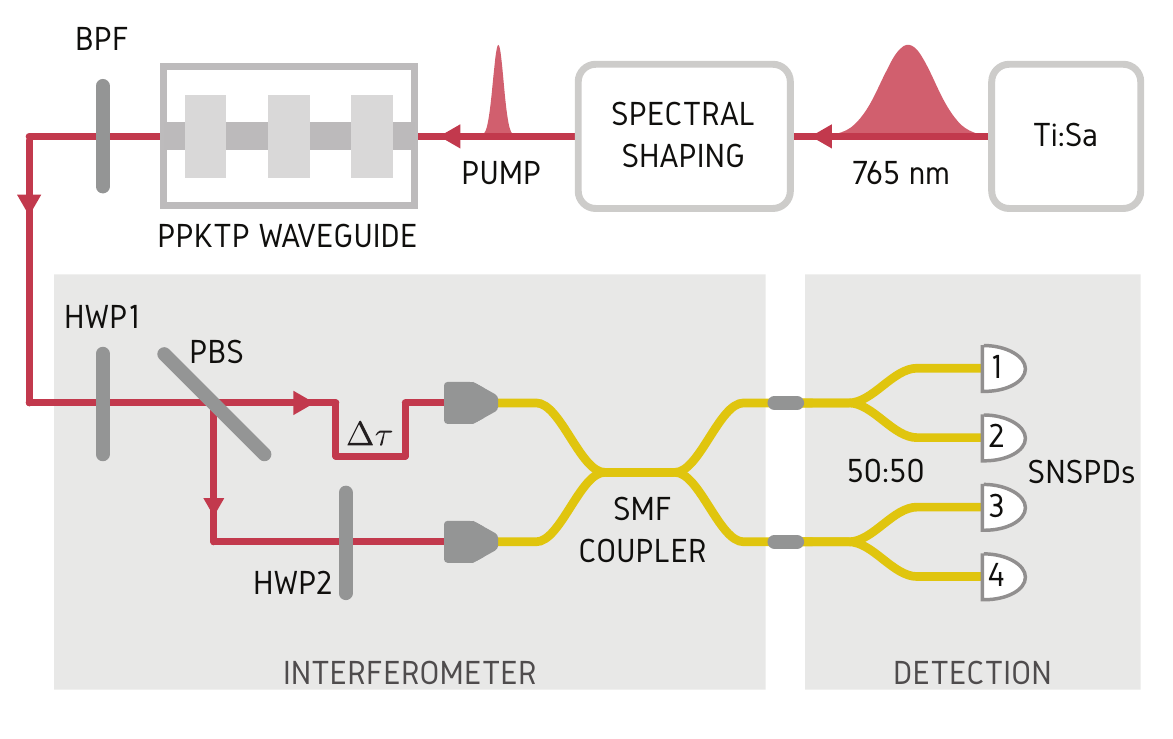}
\caption{Experimental setup. A femtosecond titanium:sapphire (Ti:Sa) oscillator with repetition rate of 80 MHz is used to pump a PPKTP waveguide designed for type-II PDC. For spectral shaping of the pump, we use a spatial light modulator (SLM) in a folded 4f setup to shape the desired spectral amplitude and phase. An 8 nm wide bandpass filter (BPF) centered at 1532 nm was used to block the pump and phasematching side-lobes. The orthogonally polarized PDC photons were sent to the interferometer setup where we used a polarising beamsplitter (PBS), a half waveplate (HWP), and an adjustable time delay stage $\Delta\tau$ to control the interference. Then the photons were sent to a single-mode fibre coupler with an adjustable coupling ratio where interference happens. Each output port of the fibre coupler is then connected to a balanced fibre splitter followed by superconducting nanowire single photon detectors (SNSPD).}
\label{exsetup}
\end{figure}

To study the impact of time-frequency correlations of the state on the four-photon interference, we employ an engineered, programmable PDC source. Our source is an 8 mm long waveguided channel in KTP, engineered to the symmetric group velocity matching condition which allows us to flexibly control the frequency correlation between signal and idler photons by modulating the pump pulses only \cite{harder2013optimized, ansari2018tailoring}. We pump the source with ultrashort pulses out of an Ti:Sa oscillator, followed by a pulse shaper. The pulse shaper is based on a spatial light modulator in a 4f setup which allows us to shape the spectral amplitude and phase of pump pulses. With this configuration we can have pulses with a time bandwidths ranging from 0.3 to 40 picoseconds. Further details of the experimental setup are given in Fig. \ref{exsetup} \cite{harder2013optimized, ansari2018tomography}.

In this work we consider three PDC states (A, B, C), generated by pump pulses with bandwidths of 0.14, 1.29, and 6.62 picoseconds, as summarized in Table \ref{tab:statesoverview}. To shape the temporal profile of the pump field in the case of states A and B we simply increase the pulse duration and carve out the corresponding spectral amplitude with a constant spectral phase. This approach works pretty well for short pulse duration but for longer pulse duration results in a significantly small pulse energies which in turn reduces the probability of generating PDC photons. For this reason, in the case of state C we take an alternative method; we use the same spectral amplitude as in the case A but with a quadratic spectral phase with the constant $D=1.9 \ \mathrm{ps}^2$. This additional phase modifies the JSA as follows
\begin{equation}
\bar F(\omega_s,\omega_i)=F(\omega_s,\omega_i)e^{i D (\omega_s+\omega_i-\omega_p)^2}.
\label{chirped}
\end{equation}

\begin{table}[t]
\centering
\caption{Overview of studied PDC states. All bandwidths $\Delta$ are referring to the standard deviation of the amplitude the optical field.}
\begin{tabular}{llll}
PDC state & A & B & C \\
\hline
$\Delta \lambda_{\mathrm{pump}}$ (nm) 	& 1.8 \ \ 				& 0.2 \ \ 				& 1.8 \\
						
$\Delta \omega_{\mathrm{pump}}$ (THz) 	& 3.479 \ \ 			& 0.386 \ \ 			& 3.44 \\
$D$ ($\mathrm{ps}^2$) 					& 0 \ \ 				& 0 \ \ 				& 1.9 \\
$\Delta t_{\mathrm{pump}}$ (ps)			& 0.14 					& 1.29 \ \ 				& 6.62 \\
$\textsl{g}^{(2)}(\tau =0)$ 			& $1.897\pm0.011$ \ \ 	& $1.233\pm0.010$ \ \ 	& $1.108 \pm 0.003$ \\

$K=\frac{1}{\textsl{g}^{(2)}-1}$ 		& 1.11 					& 4.29 \ \ 				& 9.25 \\

\hline
\end{tabular}
  \label{tab:statesoverview}
\end{table}

\begin{figure*}[ht]
\includegraphics[width=1\linewidth]{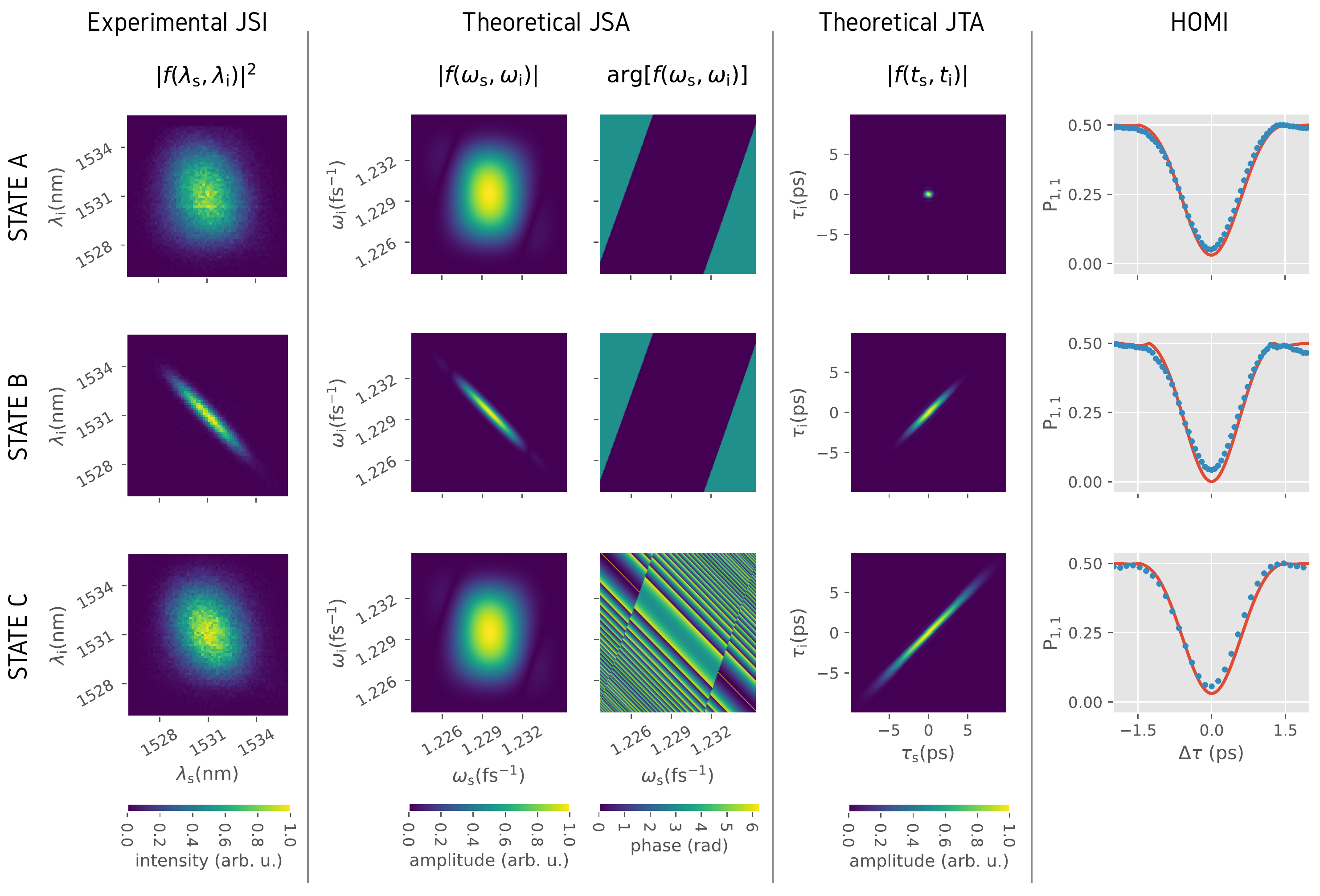}
\caption{Spectral-temporal properties of considered PDC states: state A is a nearly decorrelated PDC, state B is a standard frequency anti-correlated PDC, and state C is a PDC state with spectral phase anti-correlations from a strongly chirped pump. The first column presents measured joint spectral intensity (JSI) which contains no information about the spectral phase. The second column depicts the absolute value and the phase of theoretical joint spectral amplitudes (JSAs). The third column is the absolute value of theoretical joint temporal amplitudes (JTAs). The fourth column shows the calculated (red solid line) and measured (blue dots) two-photon Hong-Ou-Mandel intereference (HOMI), with error bars smaller than the dots.}
\label{fig:jsi}
\end{figure*}

To characterize the joint spectral intensity (JSI), $|F (\omega_s, \omega_i)|^2$, of the PDC states, we employ a  time-of-flight spectrometer with a resolution of 0.1 nm \cite{avenhaus2009fiber}. In the first column of Fig. {\ref{fig:jsi}} we plot measured JSIs of three PDC states. Since the JSI does not contain any information about the spectral phase, the JSI of the state A and C are essentially identical. State B, however, features a strong frequency anti-correlation. The second column of Fig. \ref{fig:jsi} presents theoretically calculated spectral amplitude and phase of considered JSAs; it is clearly seen that states A and C are completely different due to the phase.

\begin{figure*}
\includegraphics[width=1\linewidth]{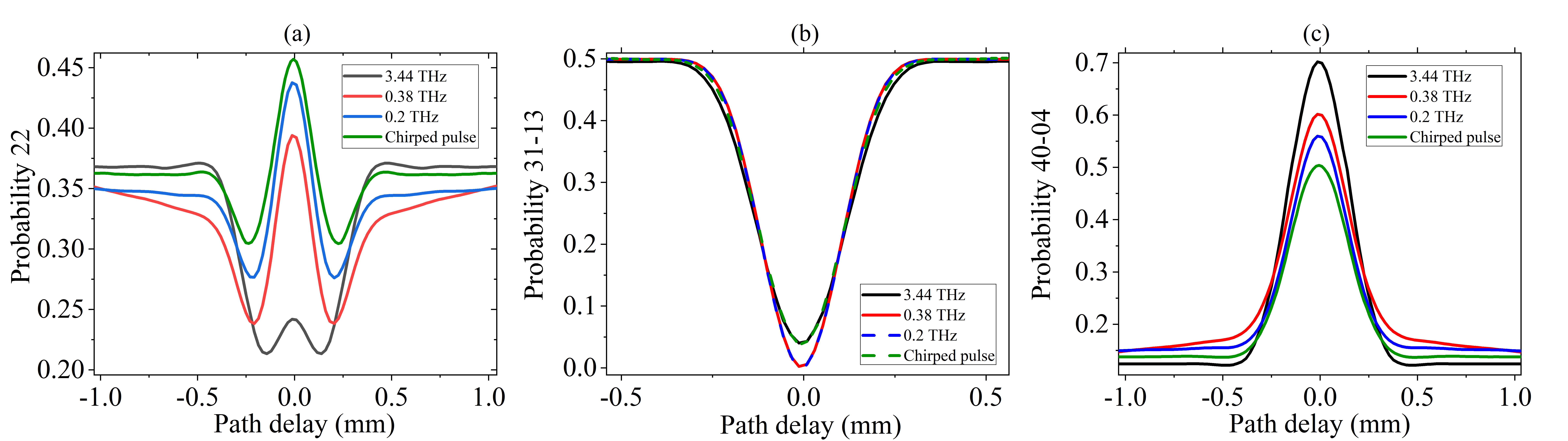}	
\caption{Theoretical coincidence probabilities to detect: (a) two photons per channel, (b)  three photons in one channel and one photon in the other channel (the green and blue curves are dotted in order to show the overlapping with the black and red curves respectively), (c) four photons in one channel for different pump pulses.}
\label{balpulse}
\end{figure*}
\begin{figure*}
\includegraphics[width=1\linewidth]{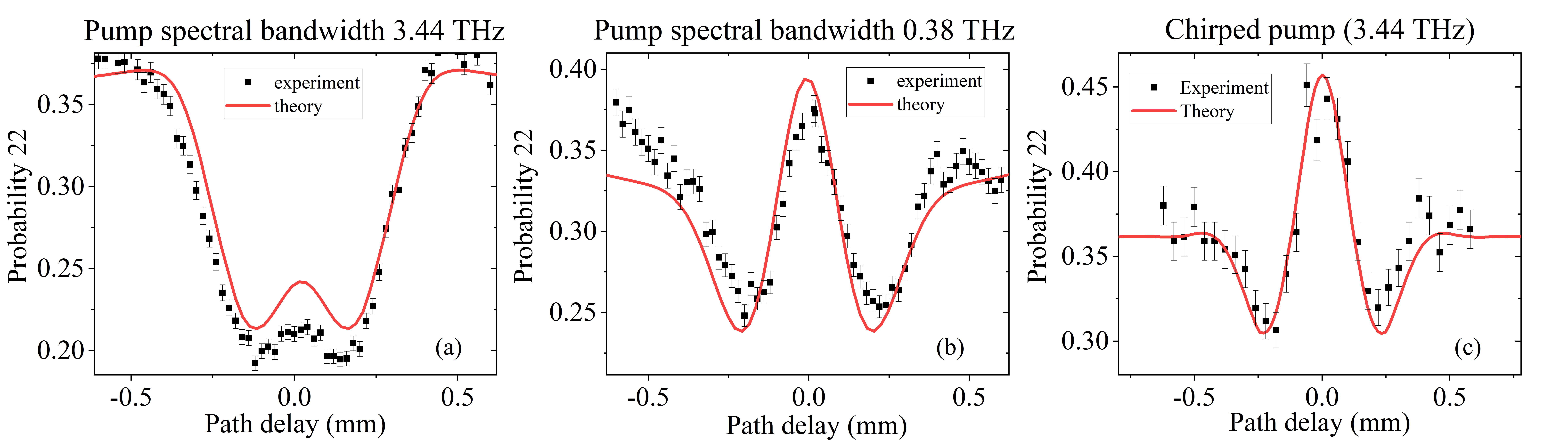}
\caption{ Comparison between theory and experiment of the $P_{22}$ probability in cases a) A, b) B and c) C.}
\label{exp1}
\end{figure*}

To find a better understanding of our experiment, it would be helpful to translate the spectral representation to temporal description by means of Fourier transform,  since the HOM interference shows an indistinguishability of photons with the same polarization and mode structure in time
\begin{equation}
 F(t_s,t_i)= \int  \mathsf{d}\omega_s  \mathsf{d}\omega_i F(\omega_s,\omega_i)e^{i (\omega_s t_s+\omega_i t_i)}.
\label{Fourier}
\end{equation}
The absolute value of the joint time amplitude (JTA), Eq. (\ref{Fourier}), is depicted in the third column of Fig. \ref{fig:jsi}. In these plots, it is clearly seen that the temporal correlations between signal and idler photons, the multimodeness in the state, is increasing from state A to B and to C. Despite distinctively different pump pulse bandwidths, the width of the two-photon HOM interference is mainly determined by the fixed crystal length and thus nearly identical for all three states, as experimentally shown in fourth column of {Fig. \ref{fig:jsi}}. This independence of the two-photon HOM interference on pump pulse duration  is a consequence of special dispersion characteristic of our PDC source which has been reported in the past  {\cite{kuzucu2005two, ansari2014probing}}. 

 The visibility of the HOM dip, on the other hand, is governed by the symmetry of JSA around the main diagonal $\lambda_s = \lambda_i$: for the non-symmetrical JSA even for a zero time delay the probability to observe two photons in two channels is not zero.

To characterize the Schmidt number $K=\frac{1}{\textsl{g}^{(2)}-1}$ of PDC states, we measure the unheralded second-order correlation function $\textsl{g}^{(2)}(\tau =0)$ \cite{christ2011probing}. The $\textsl{g}^{(2)}$ measurement probes the photon number statistics of signal or idler photons, where a single-mode PDC state shows a $\textsl{g}^{(2)}=2$ and a highly multimode state gives $\textsl{g}^{(2)}=1$. As shown in Table \ref{tab:statesoverview}, state A is nearly single-mode while states B and C have an increasing amount of multimodeness.

\subsection{Balanced BS}
In this section, we investigate the four-photon interference using the aforementioned three PDC states. In the first part of our analysis, we use  a balanced BS, i.e. having the same values for the reflection and transmission coefficients. The probabilities $P_{mn}$ to detect $m$  photons in the upper channel and $n$ photons in the lower channel described by Eq. (\ref{P22},\ref{P31},\ref{P40}) are plotted in Fig.\ref{balpulse}.

We can clearly observe that by varying the pump pulse duration the four-photon HOM interference is modified considerably, unlike what was expected in the case of two-photon interference, seen in Fig. \ref{fig:jsi}. Such behaviour reflects the complexity of the multiphoton interference and can be explained by the number of Schmidt modes \cite{law2000continuous} in the PDC state.  Indeed, the black line in Fig.\ref{balpulse} corresponds to the quasi-single-mode case, state A, with the Schmidt number $K=1.1$ and near circular JTA. Increasing the pump temporal duration, states B and C, results in a higher number of modes which leads to a stronger anti-bunching effect at zero time delay which is reflected in growing of the $P_{22}$ probability.

Simultaneously, for the zero time delay under the assumption of symmetrical JSA, i.e. $F(\omega_s, \omega_i) = F(\omega_i, \omega_s)$, the $P_{22}$ and $P_{40} = 1-P_{22}$ probabilities can be calculated analytically with using of the Schmidt decomposition of JSA : $F (\omega_s, \omega_i) = \sum_n  \sqrt{\Lambda_n} u_n (\omega_s) v_n (\omega_i)$, where $\Lambda_n$ are eigenvalues, $ u_n $ and $ v_n$ are eigenfunctions of the Schmidt decomposition   \cite{law2000continuous} (see Supplementary):
\begin{equation}
P_{22}=\frac{1}{2+2 \sum_n \Lambda_n^2}.
\label{chirped}
\end{equation}
\begin{figure*}[ht]
\centering
\includegraphics[width=1\linewidth]{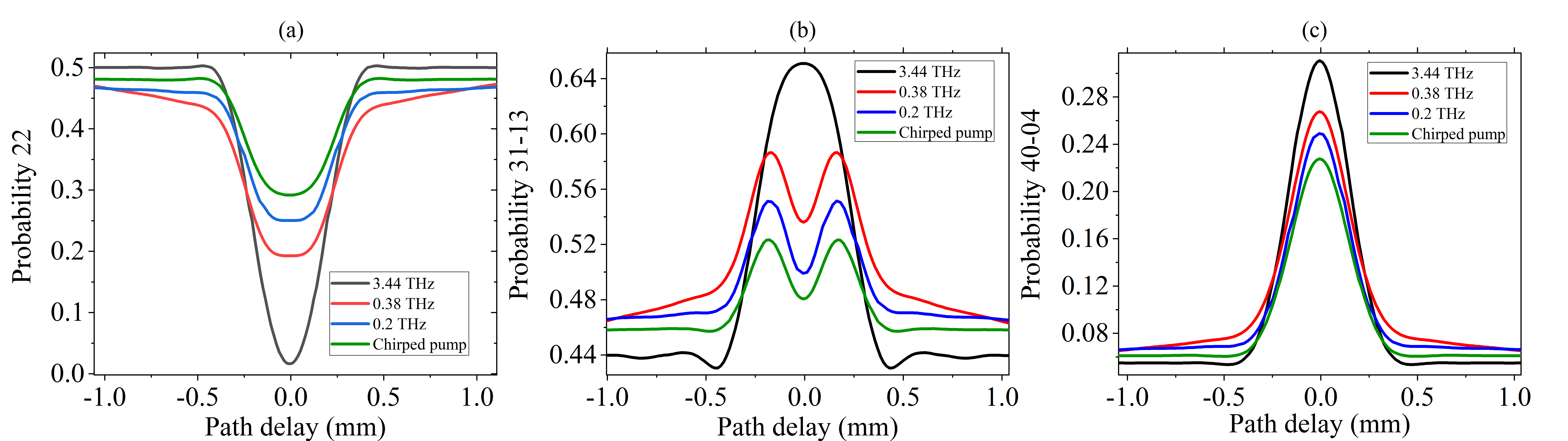}
\caption{Theoretical coincidence probabilities to detect: (a) two photons per channel, (b) three photons in one channel and one photon in the other channel, (c) four photons in one channel for different pump pulses in the unbalanced BS case.}
\label{unbalpulse}
\end{figure*}
\begin{figure*}[ht]
\includegraphics[width=1\linewidth]{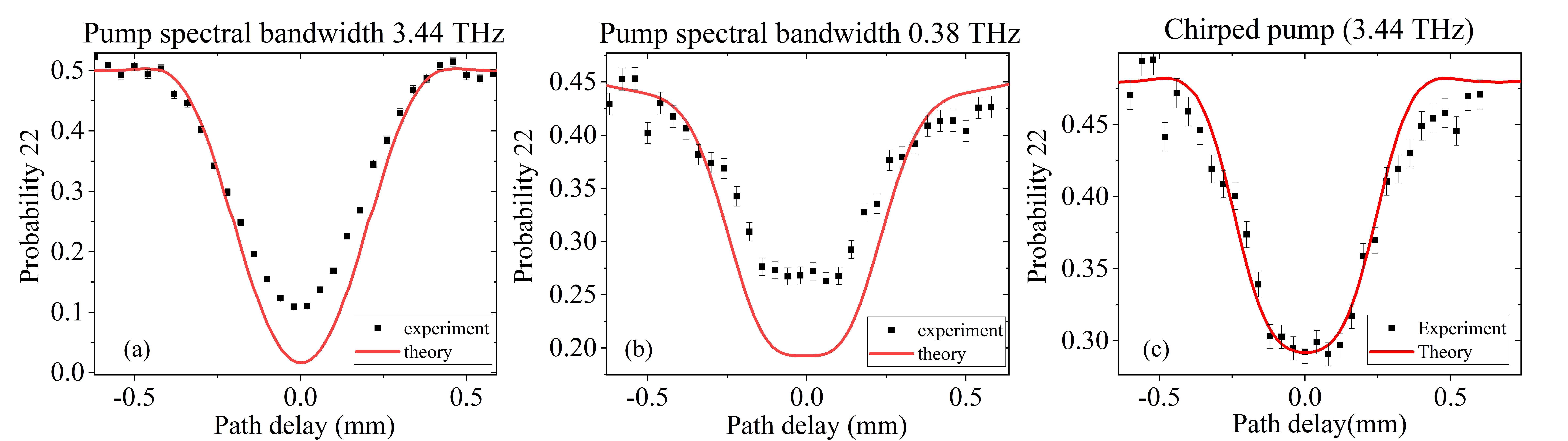}
\caption{Comparison between theory and experiment of the $P_{22}$ probability for cases a) A, b) B and c) C in the unbalanced BS configuration.}
\label{exp2}
\end{figure*}
It means that $P_{22}$ can be varied  from $P_{22} = 1/4$ in the single mode regime to the $P_{22} = 1/2$ in the strongly multimode case. This relation between the degree of multimodeness, which is characterized by the Schmidt number $K = 1/ \sum_n \Lambda_n^2 $, and antibunching properties of light can be observed only in the case of multiphoton interference, which is highly relevant for any quantum networks dealing with multimode many-particle states like boson sampling.  It is important to note that there are no analogous variations in the case of the two-photon interference.

In the case of state C the JSI is the same as in the quasi-single-mode case A but due to the quadratic phase of pump pulse, their JTAs are completely different, see Fig. \ref{fig:jsi}, the last leads to different interference patterns. However, for states of light with the same JTAs the coincidence probabilities are identical since behavior of the multiphoton interference is determined by temporal properties of light. For example, the interference patterns in the case C and in the case of the small pump spectral bandwidth $0.2$ THz are close to each other (Fig. \ref{balpulse}b, green and blue lines) and show the multimodeness  of light.

Moreover, in the four-photon case  the symmetry of JSA is very important because it is strongly  affects the $P_{31}$ probability. In particular, in cases A and C the JSA is not fully symmetrical around the main diagonal, this leads to the non-zero $P_{31}$ probability even for zero time  delay, Fig.\ref{balpulse}b, black and green curves. Nevertheless, by decreasing the pump spectral bandwidth the JSA becomes more symmetrical respect to the signal-idler variables exchange, case B, and as a consequence, $P_{31}$ vanishes when $\Delta l = 0$ (red line in Fig. \ref{balpulse}b).

To justify our theoretical analysis we built the experimental setup is sketched in Fig. \ref{exsetup}. Four detectors in the end of the setup allow to analyze the four-photon interference. The comparison between measured coincidence probabilities and theoretical calculations are illustrated in Fig.\ref{exp1} by black and red curves respectively.

\subsection{Unbalanced BS}

A balanced beam splitter, i.e. a beam splitter having the same values for both transmission and refection coefficients, is a fundamental tool in the two-photon interference scenario, since it allows to inhibit the probability to measure two photons in different channels. In the four-photon scenario, a balanced BS is able to annul the $P_{31}$ probability but cannot erase the $P_{22}$ probability. That is why it is also interesting to take into account another values of the BS parameters, which allow to inhibit the output state with two photons in both channels and provide an analogy with the two-photon interference. The values of the transmission and reflection coefficients of the BS which vanish the $P_{22}$ probability for the zero time delay in the single-mode (plane wave) case are $(3 \pm\sqrt{3})/6$ \cite{liu2007four}.

In Fig.\ref{unbalpulse} one can observe the behaviour of the $P_{mn}$ probabilities for the different JSAs depicted in Fig.\ref{fig:jsi}. As it clearly seen, with using the unbalanced BS we can inhibit drastically the $P_{22}$ by imposing a circular JSA, black curve in Fig. \ref{unbalpulse}a. Nevertheless, it is not possible to annul the $P_{22}$ totally since the JSA is not perfectly circular.  With increasing the number of modes the $P_{22}$ is growing.  It is worth to observe, that the $P_{31}$ probability, which shows a dip using a balanced BS, transforms to  the broad peak in the unbalanced case, which is consequence of the unbalanced BS. The comparison with the experimental data is shown in Fig.\ref{exp2}.

\begin{figure}[ht]
\centering
\includegraphics[width=1\linewidth]{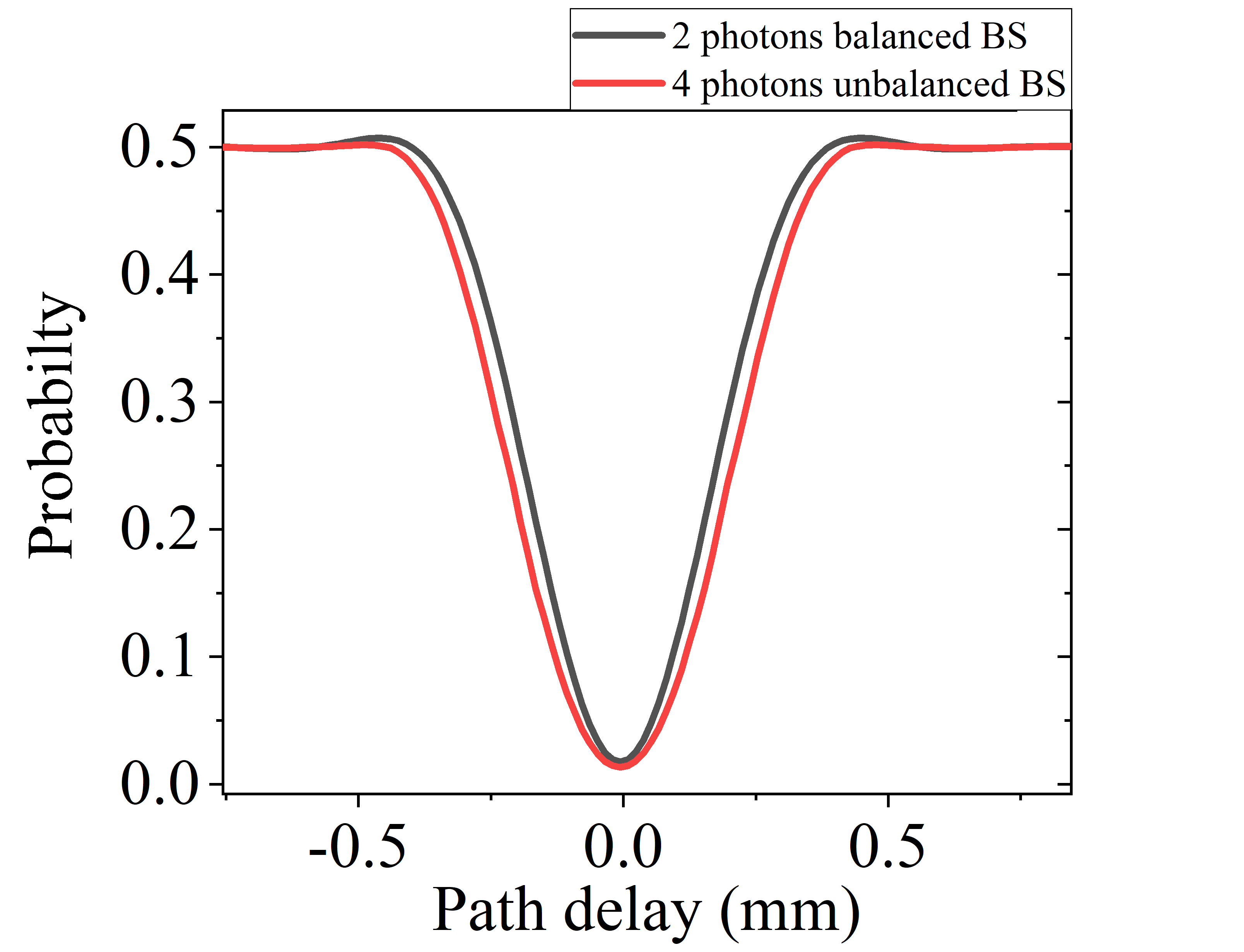}
\caption{The two-photon HOM dip with a balanced BS and the four-photons HOM dip with an unbalanced BS in the single-mode regime: the pulse duration is 0.29 ps.}
\label{2vs4}
\end{figure}
 
In the case of the unbalanced BS and taking the single-mode state, the $P_{22}$ probability is similar to the typical two-photon HOM-dip and we can therefore compare such case with the two-photon interference (the $P_{11}$ probability) by using the same parameters but the balanced BS. The comparison is presented in Fig.\ref{2vs4}, one can observe that the width of the dip in $P_{22}$ and $P_{11}$ cases are close to each other. Hovewer, the minimum point in the $P_{22}$ case is deeper respect to the $P_{11}$, which means  that $P_{22}$ is more stable against the non-symmetrical JSA. Simultaneously the $P_{22}$ is broader in comparison to the $P_{11}$ (FWHM $\approx 0.37$mm for two photons, FWHM $\approx 0.4$mm for four photons) that indicates larger range of indistinguishability of four photons due to the unbalanced BS.

\section{Conclusion}
The four-photon Hong-Ou-Mandel interference was investigated by using different mode content of the PDC source. It was observed that the HOM profiles depend strictly on the parameters of the source in terms of both the number of Schmidt modes and the symmetry of the JSA.  An antibunching behavior in the interference pattern of four-photon interference is directly connected with the number of temporal modes (multimodeness) in the system and becomes more pronounced  with increasing the Schmidt parameter. Such behaviour can be observed in the case of multiphotom interference only. 

The number of modes was modified by varying the pump spectral bandwidth and for the first time with using of a chirped pump pulse. The last case with artificial creating of multimodenes by adding a quadratic spectral phase allows to increase significantly the number of modes without decreasing the pulse energy and maintaining the same signal-idler spectrum. 

Also we demonstrated that it is possible to change drastically the shape of the HOM curves by varying the transmission and reflection parameters of the beam splitter. It was shown that with a specific choice of such parameters the four-photon interference can be similar to the two-photon interference but with larger range of photons indistinguishability and better stability to the asymmetry of JSA. 

Presented results illuminate features of  multiphoton interference with using a real multimode photon source and open an avenue for further investigation of multiphoton interfernece and its implementation into quantum networks and quantum computing algorithms based on photonic structures.

\section{ACKNOWLEDGMENTS}
We acknowledge the financial support of the Deutsche Forschungsgemeinschaft (DFG)
via TRR~142/1, project C02.  P.~R.~Sh. thanks the state of North Rhine-Westphalia for support by the
{\it Landesprogramm f{\"u}r geschlechtergerechte Hochschulen}.

\bibliography{database}

\begin{widetext}

\section{Appendix}
\subsection{Relation between $P_{22}$ and the number of modes}
In this Appendix we show how the peak in $P_{22}$ at the zero time delay depends on the number of Schmidt modes of the source. At the zero time delay, the expression for $P_{22}$ probability can be drastically reduced due to the symmetry of the JSA respect to the main diagonal:
\begin{equation}
P_{22}=\frac{\int \mathsf{d}\omega_b \mathsf{d}\omega_c \mathsf{d}\tilde\omega_b \mathsf{d}\tilde\omega_c F(\omega_c,\omega_d)F(\tilde\omega_c,\tilde\omega_d)F^*(\omega_d, \omega_c)F^*(\tilde\omega_d,\tilde\omega_c)}{2+2\int\mathsf{d}\omega_b \mathsf{d}\omega_c\mathsf{d}\tilde\omega_b \mathsf{d}\tilde\omega_c F(\omega_c,\omega_d)F(\tilde\omega_c,\tilde\omega_d)F^*(\omega_d, \tilde\omega_c)F^*(\tilde\omega_d,\omega_c)}.
\end{equation}
The integration in the numerator tends to the unity due to the normalization of the JSA. To calculate the expression in the denominator it is helpful to perform the Schmidt decomposition of the JSAs:
\begin{equation}\begin{split}
\int\mathsf{d}\omega_b \mathsf{d}\omega_c\mathsf{d}\tilde\omega_b \mathsf{d}\tilde\omega_c F(\omega_c,\omega_d)F(\tilde\omega_c,\tilde\omega_d)F^*(\omega_d, \tilde\omega_c)F^*(\tilde\omega_d,\omega_c)=\\
\sum_{\alpha\beta\gamma\delta}\sqrt{\Lambda_\alpha \Lambda_\beta\Lambda_\gamma\Lambda_\delta}\int\mathsf{d}\omega_b \mathsf{d}\omega_c\mathsf{d}\tilde\omega_b \mathsf{d}\tilde\omega_c u_\alpha(\omega_c)u_\beta(\tilde\omega_c)u^*_\gamma(\omega_d)u^*_\delta(\tilde\omega_d)v_\alpha(\omega_c)v_\beta(\tilde\omega_c)v^*_\gamma(\omega_d)v_\delta^*(\tilde\omega_d),
\end{split}\end{equation}
where parameters $\Lambda_n$ and  functions $u_n(\omega)$ and $v_n(\omega)$ are Schmidt eigenvalues and eigenfunctions respectively. Due to the symmetry of JSA we can assume $u\equiv v$. With using of the orthonormalization of the Schmidt-mode basis, the integral in the denominator can be taken and the final $P_{22}$ probability at the zero time delay can be obtained:
\begin{equation}
P_{22}=\frac{1}{2(1+\frac{1}{k})},
\end{equation}
where $K=1/(\sum_\alpha \Lambda_\alpha^2)$ is the Schmidt number. How it was predicted theoretically and confirmed experimentally, such probability at the zero time delay depends strictly on the number of Schmidt modes. When the JTA is extremely narrow, the number of Schmidt modes increases. Such situation leads to the bunching behavior observed both theoretically and experimentally  in Fig.\ref{balpulse} and Fig.\ref{exp1} respectively. By increasing the number of Schmidt modes either via increasing the pump spectral bandwidth or by using a quadratic phase chirped pump laser, the system experiences an higher degree of entanglement and as consequence of that, an antibunching peak is observed.

\end{widetext}

\end{document}